\documentclass[11pt]{article}
\usepackage{graphicx}
\usepackage{graphics}
\usepackage{subfigure}
\usepackage{amsmath}
\usepackage{amscd}
\usepackage{amssymb}
\usepackage{color}
\usepackage[all]{xy}

\newtheorem{theorem}{Theorem}[section]
\newtheorem{lemma}[theorem]{Lemma}
\newtheorem{remark}[theorem]{Remark}

\newtheorem{notation}{Notation}
\newtheorem{definition}{Definition}[section]
\newtheorem{proposition}{Proposition}[section]
\newtheorem{corollary}{Corollary}[section]
\newcommand{\qed}{\hfill\rule{0.2cm}{0.2cm}}

\newcommand{\rank}{{\mathop{\mathrm{rank}}}}
\newcommand{\Res}{{\mathop{\mathrm{Res}}}}
\newcommand{\diag}{{\mathop{\mathrm{diag}}}}
\title{Using Smith Normal Forms and $\mu$-Bases to Compute All the Singularities of Rational
Planar Curves}
\author{ Xiaohong Jia$^{ a}$,~~  Ron
  Goldman$^{b}$\\
{\small $^a$Department of Mathematics, University of Science}\\
{\small and Technology of China, Hefei, 230026}\\
{\small $^b$Computer Science Department, Rice University,}\\
{\small 6100 Main St., MS-132, Houston, TX 77005, USA} }

\date{}

\begin{document}
\maketitle

\begin{abstract}
We prove the conjecture of Chen, Wang
and Liu in \cite{8Chen03} concerning how  to calculate the parameter
values corresponding to all the singularities, including the
infinitely near singularities, of rational planar curves from the
Smith normal forms of certain Bezout resultant matrices derived from
$\mu$-bases.
\end{abstract}

\vskip 1cm


\section{Introduction}
The nature and number of the singularities of planar algebraic
curves contain a great deal of information about the geometry and
topology of these curves. Therefore much has been written about how
to compute these singularities \cite{Chen and
Sederberg(2002),Coorlidge(1931),Fulton(1989),Hilton(1920),Perez-Diaz(2007),Walker(1950)}.
Recently, Chen, Wang, and Liu presented a conjecture concerning how
to use the Smith normal forms of certain Bezout resultant matrices
derived from $\mu$-bases to calculate the parameter values
corresponding to all the singularities, including the infinitely
near singularities, of rational planar curves \cite{8Chen03}. The
goal of this paper is to prove their
conjecture.

This paper is a sequel to our paper on {\it $\mu$-Bases and
Singularities of Rational Planar Curves }\cite{JiaGoldman}, where we
show how to compute the parameters corresponding to each
singularity, including the infinitely near singularities, of
rational planar curves using certain Bezout resultant matrices
derived from $\mu$-bases. Here we shall show how to use the approach
established in that paper to prove  the
conjecture of Chen, Wang, and Liu.

 We proceed in the
following fashion. In Section 2, we review the notions of
$\mu$-bases for rational planar curves and Smith normal forms of
polynomial matrices. In Section 3 we summarize the main results in
\cite{JiaGoldman} on the computation and analysis of the
singularities, together with their infinitely near singularities, of
rational planar curves. In Section 4 we state the conjecture of
Chen, Wang and Liu. Section 5 is devoted to a proof of our
main result. Here we focus on a single singularity. First we compute
the Smith normal forms of two Hybrid Bezout matrices, one of which
provides all the parameters of the infinitely near singularities
while the other provides all the parameters of the original
singularity. We then combine these two Smith normal forms together
by invoking companion matrices to factor the $k$-th determinant
factors of the Bezout matrix that appears in our main theorem and
thereby complete our proof. We close in Section 6 with a more
detailed discussion of the relationship between our main theorem and
the conjecture of Chen, Wang and Liu.

\section{Preliminaries: $\mu$-bases and Smith normal forms}

We begin by reviewing some preliminary concepts which we shall need
in the statement and proof of our main result.

 \subsection{$\mu$-bases}
 Let $\mathbb{R}[s,u]$ be the set of homogeneous polynomials in the
homogeneous parameter $s:u$ with real coefficients. A
parametrization for a degree $n$ rational planar curve is usually
written in homogeneous form  as
\begin{equation}
\mathbf{P}(s,u)=(a(s,u),b(s,u),c(s,u)), \label{para of P(s,u)}
\end{equation}
 where
$a(s,u),b(s,u),c(s,u)$  are degree $n$ homogeneous polynomials in
$\mathbb{R}[s,u]$.  To avoid the degenerate case where
$\mathbf{P}(s,u)$ parameterizes a line, we shall assume that the
three homogeneous polynomials $a(s,u),b(s,u),c(s,u)$ are relatively
prime and linearly independent. Moreover, throughout this paper we
will assume that the parametrization  $\mathbf{P}(s,u)$ is
generically one-to-one.

A polynomial vector $\mathbf{L}(s,u)=(A(s,u),B(s,u),C(s,u))$ is a
{\it syzygy} of the parametrization (\ref{para of P(s,u)}) if
\begin{equation}
\mathbf{L}(s,u)\cdot\mathbf{P}(s,u)=A(s,u)a(s,u)+B(s,u)b(s,u)+C(s,u)c(s,u)\equiv0.
\label{LcdotP}
\end{equation}
The set $\mathbf{M}_p$ of all syzygies of a rational planar curve
$\mathbf{P}(s,u)$ is a module over the ring $\mathbb{R}[s,u]$,
called the {\it syzygy module}.
 The syzygy module $\mathbf{M}_p$ is known to be a free module with two
generators \cite{5Chen-wang02}.

\begin{definition}{}
Two syzygies $\mathbf{p}(s,u)$ and $\mathbf{q}(s,u)$ are called a
$\mu$-basis for the rational planar curve $\mathbf{P}(s,u)$ if
$\mathbf{p}$ and $\mathbf{q}$ form a basis for $\mathbf{M}_p$, i.e.,
every syzygy $\mathbf{L}(s,u)\in\mathbf{M}_p$
 can be written as
\begin{equation}
\mathbf{L}(s,u)=\alpha(s,u)\mathbf{p}(s,u)+\beta(s,u)\mathbf{q}(s,u),
\end{equation}
where $\alpha(s,u),\beta(s,u)\in \mathbb{R}[s,u].$
\label{D:homo_def}
\end{definition}

Note that since we are using  homogeneous polynomials, Definition
\ref{D:homo_def} implicitly implies the following degree constraint
of the elements of a $\mu$-basis \cite{Cox}:
$$\deg(\mathbf{p})+\deg(\mathbf{q})=\deg(\mathbf{P}).$$

Every rational planar curve has a $\mu$-basis. Moreover, there is a
fast algorithm for computing $\mu$-bases based on Gaussian
elimination \cite{5Chen-wang02}.

$\mu$-bases have many advantageous properties. For example,  we can
recover the parametrization of the rational planar curve
$\mathbf{P}(s,u)$ from the outer product of a $\mu$-basis:
\begin{equation}
\mathbf{p}(s,u)\times\mathbf{q}(s,u)=k\mathbf{P}(s,u),
\label{crossproduct}
\end{equation}
 where $k$ is a nonzero constant. We can also retrieve
the implicit equation $f(x,y,w)=0$ of the rational planar curve
$\mathbf{P}(s,u)$ by taking the resultant of a $\mu$-basis:
\begin{equation}
f(x,y,w)=\Res_{s,u}(\mathbf{p}(s,u)\cdot\mathbf{X},\mathbf{q}(s,u)\cdot\mathbf{X}),
\end{equation}
where $\mathbf{X}=(x,y,w)$ \cite{5Chen-wang02}.

 \subsection{Smith normal
  forms}
  The statement and proof of our main results concern matrices whose
  entries are polynomials. To study these
  matrices, we are going to employ Smith normal forms. The definition and
  main properties of Smith normal forms are reviewed below and summarized in Definitions \ref{invertibledef}--\ref{equivdef}
  and Propositions \ref{invertible}--\ref{equivalentsmithform}; for further
  details and proofs, see \cite{La}.

\begin{definition}
A polynomial matrix $P\in M_{m\times m}(\mathbb{R}[t])$ is said to
be invertible if  $\det(P)=c\in \mathbb{R}$ and $c\not=0$.
\label{invertibledef}
\end{definition}

\begin{proposition}
The following elementary row matrices in $M_{m\times
m}(\mathbb{R}[t])$ are invertible:
\begin{enumerate}
  \item $E_{ij}$: interchange rows $i$ and $j$ of the identity matrix
  $I_m$;
  \item $E_i(\lambda)$: multiply row $i$ of $I_m$ by $\lambda\in \mathbb{R}, \lambda\not=0$;
  \item $E_{ij}(f)$: add $f$ times row $j$ of $I_m$ to row $i$,
  $f\in \mathbb{R}[t]$.
\end{enumerate}
Similarly the elementary column matrices
$F_{ij},~F_{i}(\lambda),~F_{ij}(f)$ in $M_{m\times
m}(\mathbb{R}[t])$ are invertible. \label{invertible}
\end{proposition}

\begin{proposition}
Each invertible polynomial matrix $P\in M_{m\times
m}(\mathbb{R}[t])$ is a product of elementary matrices.
\end{proposition}

\begin{proposition}
For every nonzero polynomial matrix $A\in M_{m\times
m}(\mathbb{R}[t])$ with $r=\rank(A)$, there exist invertible
polynomial matrices $P,Q\in M_{m\times m}(\\\mathbb{R}[t])$ such
that
\begin{equation}
PAQ=\left(
        \begin{array}{ccccccc}
          \alpha_1 &  &  &  &  &  &\\
           & \alpha_2 & &  &  & & \\
           &  & \ddots &  &  &  &\\
           &  &  & \alpha_r &  &  &\\
           &  &  &  &0  & & \\
           &  &  &  &  &\ddots & \\
           &  &  &  &  & &0 \\
        \end{array}
      \right),
\label{Smith}
\end{equation}
where $\alpha_1,\cdots,\alpha_r\in \mathbb{R}[t]$ are polynomials with
$\alpha_k|\alpha_{k+1}$ for $1\leq k<r$.
\end{proposition}

\begin{definition}
The matrix in (\ref{Smith}) is called the Smith normal form of the
polynomial matrix $A$. We shall denote the Smith normal form of $A$
by $S(A)$. Note that Smith normal forms of polynomial matrices are
unique up to constant multiples of the entries.\label{smithformdefi}
\end{definition}

\begin{definition}
Let $A,B\in M_{m\times m}(\mathbb{R}[t])$. Then A is said to be
equivalent to B over $\mathbb{R}[t]$ if and only if there are
invertible matrices $P,Q\in M_{m\times m}(\mathbb{R}[t])$ such that
$PAQ=B$. \label{equivdef}
\end{definition}

\begin{proposition}
Equivalent matrices $A,B\in M_{m\times m}(\mathbb{R}[t])$ have the
same Smith normal forms. \label{equivalentsmithform}
\end{proposition}

\begin{proposition}\cite{Robert C}
Let $A,B\in M_{m\times m}(\mathbb{R}[t])$ be nonsingular matrices, and denote by $\alpha_k, \beta_k, \gamma_k$ the $k$-th invariant factor of $A$, $B$ and $AB$, respectively. Then
$$\alpha_{i_1}\alpha_{i_2}\cdots\alpha_{i_k}\beta_{j_1}\beta_{j_2}\cdots\beta_{j_k}|\gamma_{i_1+j_1-1}\gamma_{i_2+j_2-2}\cdots\gamma_{i_k+j_k-k},$$
where the integer subscripts satisfy
$$1\leq i_1<i_2<\cdots<i_k,~~1\leq j_1<j_2<\cdots<j_k,~i_k+j_k\leq k+m.$$
\label{P:invariants}
\end{proposition}

In the proof of our main result in Section 5, we shall need the
following lemma.
\begin{lemma}
Let $F(s,t),G(s,t)$ be two bivariate polynomials of the same degree
$m$ in $t$. Then  the Smith normal form of the Bezout matrix
$B_t((s-t)F,(s-t)G)$ is
$$\diag(\alpha_1(s),\cdots,\alpha_{m-1}(s),0)$$
if and only if the Smith normal form of the Bezout matrix $B_t(F,G)$
is
$$\diag(\alpha_1(s),\cdots,\alpha_{m-1}(s)).$$
\label{sameinvariantlemma}
\end{lemma}

\noindent \textbf{Proof.} Since the two polynomials
$(s-t)F(s,t),(s-t)G(s,t)$ are of the same degree  in $t$, we can
just consider the symmetric Bezout matrix, which is the coefficient
matrix of a Bezoutian. Thus
\begin{equation}
\begin{aligned}
&(1,t,\cdots,t^{m})B_t((s-t)F,(s-t)G)(1,\alpha,\cdots,\alpha^{m})^T\\
=&\frac{\left|
          \begin{array}{cc}
            (s-t)F(s,t) & (s-t)G(s,t) \\
            (s-\alpha)F(s,\alpha) & (s-\alpha)G(s,\alpha) \\
          \end{array}
        \right|
}{t-\alpha}\\
=&(s-t)(s-\alpha)\frac{\left|
          \begin{array}{cc}
            F(s,t) & G(s,t) \\
            F(s,\alpha) & G(s,\alpha) \\
          \end{array}
        \right|
}{t-\alpha}\\
=&(s-t)(s-\alpha)(1,t,\cdots,t^{m-1})B_t(F,G)(1,\alpha,\cdots,\alpha^{m-1})^T.
\end{aligned}
\label{(s-t)Bezout}
\end{equation}
Let $q_{ij}$ be the entries in the matrix $B_t((s-t)F,(s-t)G),
i,j=1,\cdots,m+1$, and let $p_{ij}$ be the entries in the matrix
$B_t(F,G), i,j=1,\cdots, m$. Also set $p_{i,m+1}=p_{m+1,j}=0$ for
any $i,j=1,\cdots,m+1$. Then from (\ref{(s-t)Bezout}), for
$i,j=2,\cdots,m+1$
\begin{equation}
q_{ij}=s^2p_{ij}-s(p_{i,j-1}+p_{i-1,j})+p_{i-1,j-1}.
\label{rowoperation}
\end{equation}
The right hand side of equation (\ref{rowoperation}) represents row
and column operations on the matrix
$$\left(
   \begin{array}{cc}
     0 & 0 \\
     0 & B_t(F,G) \\
   \end{array}
 \right).
$$
But by Proposition \ref{equivalentsmithform} row and column
operations do not alter the Smith normal form because by Proposition
\ref{invertible} elementary matrices are invertible.
 Therefore, the Smith normal form of the
Bezout matrix $B_t((s-t)F,(s-t)G)$ is
$$\left(
    \begin{array}{cc}
      S(B_t(F,G)) &  \\
       & 0 \\
    \end{array}
  \right),
$$where $S(B_t(F,G))$ is the Smith normal form of the Bezout matrix
$B_t(F,G)$. \qed

\medskip

Definition \ref{smithformdefi} deals with Smith normal forms of
univariate polynomial matrices. But in our work on rational curves
we deal mainly with homogeneous polynomials. Therefore next we
provide a definition for Smith normal forms of  matrices whose
entries are homogeneous polynomials.
\medskip

\begin{definition}
Let $A(t,v)\in M_{m\times m}(\mathbb{R}[t,v])$. Suppose that the
Smith normal form of the matrix $A(t,1)$ is
$$\diag(\bar{d}_m(t), \bar{d}_m(t)\bar{d}_{m-1}(t),\cdots,\bar{d}_m(t)\cdots\bar{d}_1(t)),$$
and the Smith normal form of the matrix $A(1,v)$ is
$$\diag(\hat{d}_m(v),\hat{d}_m(v)\hat{d}_{m-1}(v),\cdots,\hat{d}_m(v)\cdots\hat{d}_1(v)).$$
Define
\begin{equation}
d_i(t,v)\triangleq LCM(\bar{d}_i(t,v),\hat{d}_i(t,v)),
\end{equation}
where $\bar{d}_i(t,v)$ and $\hat{d}_i(t,v)$ are the homogenizations
of the polynomials  $\bar{d}_i(t)$ and $\hat{d}_i(v)$. Then the
Smith normal form of the matrix $A(t,v)$ is given by
$$\diag(d_m(t,v),d_m(t,v)d_{m-1}(t,v),\cdots,d_m(t,v)\cdots d_1(t,v)).$$
\label{smithof homogeneous}
\end{definition}

The $k$-th determinant factor of a polynomial matrix $A$ is the GCD
of the $k\times k$ minors of the matrix $A$. The following result is
proved in \cite{8Chen03}.

\begin{proposition}\cite{8Chen03} Suppose that $A(t,v)\in M_{m\times
m}(\mathbb{R}[t,v])$. Let $D_k(t,v)$ be the determinant factors of
$A(t,v)$, and let $d_k(t,v)$ be defined as in Definition
\ref{smithof homogeneous}. Then
\begin{equation}
D_k(t,v)=d_m(t,v)^kd_{m-1}(t,v)^{k-1}\cdots
d_{m-k+1}(t,v)^2d_{m-k+1}(t,v).
\end{equation}
\label{D-khomo}
\end{proposition}

\medskip

Proposition \ref{D-khomo} implies that  the following definition is
equivalent to Definition \ref{smithof homogeneous}.

\begin{definition}
Suppose that $A(t,v)\in M_{m\times m}(\mathbb{R}[t,v])$. Let
$D_k(t,v)$ denote the determinant factors of $A(t,v)$, and let
$\alpha_i=D_{i}/D_{i-1},~i=2,\cdots,m$ and $\alpha_1=D_1$. Then the the
Smith normal form of $A(t,v)$ is given by
$$\diag(\alpha_1,\cdots,\alpha_m).$$
\end{definition}

Later we shall use the following property of Smith normal forms.
\begin{corollary}
Let $F(s,u;t,v),G(s,u;t,v)$ be two bihomogeneous polynomials of the
same degree $m$ in $t,v$. Then  the Smith normal form of the Bezout
matrix $B_{t,v}((sv-tu)F,(sv-tu)G)$ is
$$\diag(\alpha_1(s,u),\cdots,\alpha_{m-1}(s,u),0)$$
if and only if the Smith normal form of the Bezout matrix
$B_{t,v}(F,G)$ is
$$\diag(\alpha_1(s,u),\cdots,\alpha_{m-1}(s,u)).$$
 \label{sameinvariant}
\end{corollary}
\noindent Proof. Let $$B_1(s,u)\triangleq
B_{t,v}((sv-tu)F,(sv-tu)G),~~~B_2(s,u)\triangleq B_{t,v}(F,G).$$ By
Lemma \ref{sameinvariantlemma},
$$S(B_1(s,1))=\diag(S(B_2(s,1)),0),~~S(B_1(1,u))=\diag(S(B_2(1,u)),0).$$
Therefore by Definition \ref{smithof homogeneous}
$$S(B_1(s,u))=\diag(S(B_2(s,u)),0).$$\qed

\section{Previous results on singularities of rational planar curves}
In this section we review some results on the singularities of
rational planar curves derived in \cite{JiaGoldman}.

\medskip

Let $\mathbf{P}(t,v)$ be a rational planar curve with a $\mu$-basis
$\mathbf{p}(t,v),\mathbf{q}(t,v)$, and define
\begin{equation}
\begin{aligned}
&F(s,u;t,v)\triangleq \frac{\mathbf{p}(s,u)\cdot
\mathbf{P}(t,v)}{sv-tu}\\
&G(s,u;t,v)\triangleq \frac{\mathbf{q}(s,u)\cdot
\mathbf{P}(t,v)}{sv-tu}.
\end{aligned}
\label{eq_FG}
\end{equation}
Notice that $F(s,u;t,v)$ and $G(s,u;t,v)$ are polynomials, since
$$\mathbf{p}(t,v)\cdot\mathbf{P}(t,v)=\mathbf{q}(t,v)\cdot\mathbf{P}(t,v)\equiv0.$$

\begin{proposition}\cite{JiaGoldman}
A parameter pair $(s^*,u^*;t^*,v^*)$ is a common root of
$F(s,u;t,v)$ and $G(s,u;t,v)$ if and only if the two parameters
$(s^*,u^*)$ and $(t^*,v^*)$ correspond to the same singularity on
the curve $\mathbf{P}(t,v)$. \label{P:symmetric}
\end{proposition}

\begin{notation}
Let $\mathbf{Q}$ be a singular point on the rational planar curve
$\mathbf{P}(s,u)$, and let $(s_i,u_i), i=1,\cdots,k$ be all the
distinct parameters corresponding to $\mathbf{Q}$. We denote the
intersection multiplicity of $F(s,u;t,v)=0$ and $G(s,u;t,v)=0$ at
the singularity $\mathbf{Q}$ by
$$I_{\mathbf{Q}}(F,G)\triangleq \sum\limits_{i,j}I_{S_{ij}}(F,G),$$
where $S_{ij}=(s_i,u_i;s_j,u_j),$ and $I_{S_{ij}}(F,G)$ is the
intersection multiplicity of the two curves $F(s,u;t,v)=0$ and
$G(s,u;t,v)=0$ at the parameter pair $(s_i,u_i;s_j,u_j)$.
\end{notation}

\begin{proposition}\cite{JiaGoldman}
Let $\nu_{\mathbf{Q}^*}$ denote the multiplicity of an infinitely
near point $\mathbf{Q}^*$ of a singularity $\mathbf{Q}$ on the curve
$\mathbf{P}(t,v)$. Then
$$I_{\mathbf{Q}}(F,G)=\sum\limits_{\mathbf{Q}^*}\nu_{\mathbf{Q}^*}(\nu_{\mathbf{Q}^*}-1),$$
where the sum is taken over all the infinitely near points
$\mathbf{Q}^*$ of the point $\mathbf{Q}$ including $\mathbf{Q}$
itself. \label{intesectionmultiplicity for FG}
\end{proposition}


Let $\widetilde{\mathbf{p}}(s,u)$ and $\widetilde{\mathbf{q}}(s,u)$
be any pair of syzygies of the rational curve $\mathbf{P}(s,u)$ that
are linearly independent for any parameters corresponding to the
point $\mathbf{Q}$, and define
\begin{equation}
\begin{aligned}
&\widetilde{F}(s,u;t,v)\triangleq
\frac{\widetilde{\mathbf{p}}(s,u)\cdot\mathbf{P}(t,v)}{sv-tu}\\
&\widetilde{G}(s,u;t,v)\triangleq
\frac{\tilde{\mathbf{q}}(s,u)\cdot\mathbf{P}(t,v)}{sv-tu}.
\end{aligned}
\label{FGnew}
\end{equation}
Then the intersection multiplicity at $\mathbf{Q}$ of
$\widetilde{F}$ and $\widetilde{G}$ is the same as the intersection
multiplicity at $\mathbf{Q}$ of $F$ and $G$.

\begin{proposition}\cite{JiaGoldman}
$$I_{\mathbf{Q}}(\widetilde{F},\tilde{G})=I_{\mathbf{Q}}(F,G).$$
\label{changesyzygy}
\end{proposition}

In \cite{JiaGoldman}, in order to study the intersections of the two
algebraic curves $F(s,u;t,v)=0$ and $G(s,u;t,v)=0$, we focus on one
singularity $\mathbf{Q}$ of the rational planar curve
$\mathbf{P}(t,v)$. We then move the point $\mathbf{Q}$ to the origin
$(0,0,1)$ so that the parametrization of the curve $\mathbf{P}(t,v)$
has the form
\begin{equation}
\mathbf{P}(t,v)=(a(t,v)h(t,v),b(t,v)h(t,v),c(t,v)),\label{newparametrization}
\end{equation}
where $\gcd(a,b)=\gcd(h,c)=1$ and $h(t,v)$ is the inversion formula
for the singular point $\mathbf{Q}$. That is, the roots of $h(t,v)$
provide all the parameter values with proper multiplicity
corresponding to the singularity $\mathbf{Q}$. From a pair of
obvious  syzygies
\begin{equation}
\begin{aligned}
\mathbf{M}(s,u)\triangleq (-b,a,0), ~\mathbf{L}(s,u)=(c,0,-ah)
\end{aligned}
\end{equation}
we construct two additional polynomials
\begin{equation}
\begin{aligned}
M(s,u;t,v)&\triangleq
\frac{\mathbf{M}(s,u)\cdot\mathbf{P}(t,v)}{sv-tu}\\
&=\frac{a(s,u)b(t,v)-b(s,u)a(t,v)}{sv-tu}h(t,v)\\
&\triangleq \overline{M}(s,u;t,v)h(t,v)\\
L(s,u;t,v)&\triangleq
\frac{\mathbf{L}(s,u)\cdot\mathbf{P}(t,v)}{sv-tu}\\
&=\frac{c(s,u)a(t,v)h(t,v)-c(t,v)a(s,u)h(s,u)}{sv-tu}.
\end{aligned}
\label{eq_ML}
\end{equation}

By Proposition \ref{changesyzygy}, in order to examine
$I_{\mathbf{Q}}(F,G)$, we can turn to $I_{\mathbf{Q}}(M,L)$. But
$I_{\mathbf{Q}}(M,L)$ breaks into two parts:
$I_{\mathbf{Q}}(\overline{M},L)$ and $I_{\mathbf{Q}}(h,L)$, where
$I_{\mathbf{Q}}(h,L)$ gives all the parameters for the original
singular point $\mathbf{Q}$ while $I_{\mathbf{Q}}(\overline{M},L)$
gives all the parameters corresponding to the infinitely near
singularities of the point $\mathbf{Q}$. Indeed we have the
following results.

\begin{proposition}\cite{JiaGoldman}
Let $r$ be the order of the singularity $\mathbf{Q}$. Then
$$I_{\mathbf{Q}}(h,L)=r(r-1).$$
\label{intersectionforh}
\end{proposition}
\begin{proposition}\cite{JiaGoldman}
Let $\nu^{\mathbf{Q}^*}$ denote the order of the infinitely near
singularity $\mathbf{Q}^*$ of $\mathbf{Q}$. Then
$$I_{\mathbf{Q}}(\overline{M},L)=\sum\limits_{\mathbf{Q^*}}\nu^{\mathbf{Q}^*}(\nu^{\mathbf{Q}^*}-1),$$
where the sum is taken over all the infinitely near singularities
$\mathbf{Q}^*$ of $\mathbf{Q}$ not including $\mathbf{Q}$ itself.
\end{proposition}


\section{The conjecture of Chen, Wang and Liu}
In order to state the conjecture of Chen, Wang and Liu, we first
need to introduce the notion of an {\it inversion formula}.

\begin{definition}
Let $(s_i,u_i), i=1,\cdots,r$ be all the parameters corresponding to
the point $\mathbf{Q}$ on the curve $\mathbf{P}(s,u)$, i.e.,
$\mathbf{P}(s_i,u_i)=\mathbf{Q}, i=1,\cdots,r$. Then a polynomial
$h(s,u)$ whose roots are $(s_i,u_i),~i=1,\cdots,r$ is an inversion
formula for the point $\mathbf{Q}$. Similarly, for an infinitely
near singularity $\mathbf{Q}^*$ on the $k$-th blow-up curve
$\mathbf{P}^k(s,u)$, an inversion formula for the point
$\mathbf{Q}^*$ is a polynomial $h(s,u)$ whose roots are all the
parameters on the parametrization $\mathbf{P}^k(s,u)$ corresponding
to the point $\mathbf{Q}^*$, i.e.,
$\mathbf{P}^k(s_i,u_i)=\mathbf{Q}^*, i=1,\cdots,r$. Generally, the
inversion formula for $\mathbf{Q}^*$ must be a factor of the
inversion formula for $\mathbf{Q}$.
\end{definition}

\begin{remark}\cite{5Chen-wang02}
Let $\mathbf{Q}$ be a singularity on a rational planar curve
$\mathbf{P}(s,u)$ with a $\mu$-basis
$\mathbf{p}(s,u),\mathbf{q}(s,u)$. An inversion formula for
$\mathbf{Q}$ is given by
$$h(s,u)=\gcd(\mathbf{p}(s,u)\cdot\mathbf{Q},\mathbf{q}(s,u)\cdot\mathbf{Q}).$$
\label{R:inversionform}
\end{remark}

We are now ready to state the conjecture of Chen, Wang and Liu. Let
$\mathbf{P}(t,v)$ be a rational planar curve of degree $n$ with a
$\mu$-basis $\mathbf{p}(s,u)$, $\mathbf{q}(s,u)$, and let $B(t,v)$
be the Hybrid Bezout resultant matrix \cite{Fuhrmann} of the two
polynomials $\mathbf{p}(s,u)\cdot\mathbf{P}(t,v)$ and
$\mathbf{q}(s,u)\cdot\mathbf{P}(t,v)$ with respect to $(s,u)$.
Suppose that the Smith normal form of the matrix $B(t,v)$ is
$$\diag(d_{n-\mu}(t,v),d_{n-\mu}(t,v)d_{n-\mu-1}(t,v),\cdots,d_{n-\mu}(t,v)\cdots d_2(t,v),0).$$
 Then Chen,
Wang and Liu state the following conjecture \cite{8Chen03}.

\bigskip

\noindent \textbf{Conjecture} {\it
\begin{equation}
d_r(t,v)=h_r(t,v)\prod\limits_{i\geq r}\psi_r^i(t,v), \label{E:con}
\end{equation}
 where
$h_r(t,v)$ is the inversion formula of all the order $r$
singularities on the curve $\mathbf{P}(t,v)$, and $\psi_r^i(t,v)$ is
the inversion formula for all the order $r$ infinitely near
singularities in the neighborhood of order $i\geq r$ singular points
on $\mathbf{P}(t,v)$. \label{conjecture}}

\bigskip

By \cite{5Chen-wang02} we can turn to prove the following result, which is equivalent
to the conjecture of Chen, Wang and Liu.

\begin{theorem}
Let $B(s,u)$ be the Bezout resultant matrix of the two polynomials
$\mathbf{p}(s,u)\cdot\mathbf{P}(t,v)$ and
$\mathbf{q}(s,u)\cdot\mathbf{P}(t,v)$ with respect to $(t,v)$. Then
the Smith normal form of the matrix $B(s,u)$ is
$$\diag(1,\cdots,1,d_{n-\mu}(s,u),d_{n-\mu}(s,u)d_{n-\mu-1}(s,u),\cdots,d_{n-\mu}(s,u)\cdots d_2(s,u),0),$$
where $d_r(s,u), r=2,\cdots,n-\mu$ are defined in Equation
 (\ref{E:con}). \label{T:conjec_newversion}
\end{theorem}


Theorem \ref{T:conjec_newversion} is equivalent to the following
result.

\begin{theorem}
Let $B_{t,v}(F,G)$ be the Bezout resultant matrix of the two
polynomials
\begin{equation*}
F(s,u;t,v)=\frac{\mathbf{p}(s,u)\cdot\mathbf{P}(t,v)}{sv-tu},~~~G(s,u;t,v)=\frac{\mathbf{q}(s,u)\cdot\mathbf{P}(t,v)}{sv-tu}
\end{equation*}
with respect to $(t,v)$. Then the Smith normal form of the matrix
$B_{t,v}(F,G)$ is
$$\diag(1,\cdots,1,d_{n-\mu}(s,u),d_{n-\mu}(s,u)d_{n-\mu-1}(s,u),\cdots,d_{n-\mu}(s,u)\cdots d_2(s,u)),$$
where $d_r(s,u), r=2,\cdots,n-\mu$ are defined in Equation
 (\ref{E:con}).
 \label{T:conjec_newversionforFG}
\end{theorem}

The equivalence of Theorem \ref{T:conjec_newversion} and Theorem
\ref{T:conjec_newversionforFG} follows from Corollary
\ref{sameinvariant}. Therefore, instead of proving Theorem
\ref{T:conjec_newversion}, we shall prove Theorem
\ref{T:conjec_newversionforFG}.

\section{The proof of Theorem \ref{T:conjec_newversionforFG}}
We are going to prove Theorem \ref{T:conjec_newversionforFG} by
applying an approach similar to the analysis in \cite{JiaGoldman}.
We begin in subsection 5.1 by reducing Theorem
\ref{T:conjec_newversionforFG} to the computation of the Smith
normal form of the Bezout resultant matrix of two polynomials
constructed from a pair of syzygies of the curve. In subsection 5.2,
we decompose this Smith normal form into two Smith normal forms, one
of which provides all the parameters of the infinitely near
singularities while the other provides all the parameters of the
original singularity. Then in subsection 5.3, we introduce companion
matrices to factor the Bezout resultant matrices and finally we use
this factorization  to combine the two Smith normal forms together
in subsection 5.4 to complete the proof.

\subsection{Reducing to the Smith normal form of $B_{t,v}(M,L)$}
We are going to show that all the information in the Smith normal
form of the Bezout resultant matrix $B_{t,v}(F,G)$ is contained in
the Smith normal form of the Bezout resultant matrix $B_{t,v}(M,L)$,
where $M(s,u;t,v)$ and $L(s,u;t,v)$ are the two polynomials defined
in Equation (\ref{eq_ML}).
 Before we continue, a word about
our notation.

\begin{remark}
For a polynomial $d(s,u)$, we use $d^{\mathbf{Q}}(s,u)$ to denote
all the factors of $d(s,u)$ whose roots are parameters corresponding
to the point $\mathbf{Q}$. For example, if the inversion formula for
the point $\mathbf{Q}$ is $s^2(s+u)$, and $d(s,u)=s(s+u)^2(s-u)$,
then $d^{\mathbf{Q}}(s,u)=s(s+u)^2$. Consequently, if $A$ is a
polynomial matrix and
$$S(A)=\diag(f_1,\cdots,f_n),$$ then we shall write
$$S^{\mathbf{Q}}(A)\triangleq\diag(f_1^{\mathbf{Q}},\cdots,f_n^{\mathbf{Q}})$$
 to
denote the Smith normal form of $A$ restricted to $\mathbf{Q}$.
\end{remark}

Let $\widetilde{F},\widetilde{G}$ be the two polynomials in Equation
(\ref{FGnew}) generated from a pair of syzygies of the curve
$\mathbf{P}(t,v)$ that are always independent for any parameter
$(s,u)$ corresponding to the point $\mathbf{Q}$. Then we have the
following matrix version for Proposition \ref{changesyzygy}.
\begin{theorem}
$$S^{\mathbf{Q}}(B_{t,v}(F,G))=S^{\mathbf{Q}}(B_{t,v}(\widetilde{F},\widetilde{G})).$$
\label{muabdsyzygy}
\end{theorem}

\noindent Proof. Let the Smith normal form of the Bezout resultant
matrix $B_{t,v}(\widetilde{F},\widetilde{G})$ be
$$\diag(\widetilde{f}_1(s,u),\widetilde{f}_{n-1}(s,u)),$$
and let the Smith normal form of the Bezout resultant matrix
$B_{t,v}(F,G)$ be
$$\diag(f_1(s,u),\cdots,f_{n-1}(s,u)).$$
It suffices to prove that
$$f_i^{\mathbf{Q}}=\widetilde{f}_i^{\mathbf{Q}}.$$
Since $\widetilde{\mathbf{p}}(s,u),\widetilde{\mathbf{q}}(s,u)$ are
a pair of syzygies and $\mathbf{p}(s,u), \mathbf{q}(s,u)$ are a
$\mu$-basis, there are polynomials $\alpha(s,u), \beta(s,u),
\gamma(s,u), \delta(s,u)$ such that
\begin{equation*}
\begin{aligned}
&\widetilde{\mathbf{p}}(s,u)=\alpha(s,u)\mathbf{p}(s,u)+\beta(s,u)\mathbf{q}(s,u)\\
&\widetilde{\mathbf{q}}(s,u)=\gamma(s,u)\mathbf{p}(s,u)+\delta(s,u)\mathbf{q}(s,u).
\end{aligned}
\end{equation*}
Therefore
\begin{equation*}
\begin{aligned}
&\widetilde{F}(s,u;t,v)=\alpha(s,u)F(s,u;t,v)+\beta(s,u)G(s,u;t,v)\\
&\widetilde{G}(s,u;t,v)=\gamma(s,u)F(s,u;t,v)+\delta(s,u)G(s,u;t,v).
\end{aligned}
\end{equation*}
Also note that $\deg_{t,v}(F)=\deg_{t,v}(G)=n-1$, and
$\deg_{t,v}(\widetilde{F})=\deg_{t,v}(\widetilde{G})=n-1$, where
$n=\deg(\mathbf{P})$. So the Bezout resultant matrix
$B_{t,v}(\widetilde{F},\widetilde{G})$ is the coefficient matrix of
the following Bezoutian:
\begin{equation*}
\begin{aligned}
&\frac{\left|
        \begin{array}{cc}
         \alpha(s,u)F(s,u;t,v)+\beta(s,u)G(s,u;t,v)  &  \gamma(s,u)F(s,u;t,v)+\delta(s,u)G(s,u;t,v)\\
          \alpha(s,u)F(s,u;\bar{t},\bar{v})+\beta(s,u)G(s,u;\bar{t},\bar{v}) &  \gamma(s,u)F(s,u;\bar{t},\bar{v})+\delta(s,u)G(s,u;\bar{t},\bar{v})\\
        \end{array}
      \right|
}{t\bar{v}-\bar{t}v}\\
&=\frac{\left|
          \begin{array}{cc}
            F(s,u;t,v) & G(s,u;t,v) \\
            F(s,u;\bar{t},\bar{v})  & G(s,u;\bar{t},\bar{v}) \\
          \end{array}
        \right|\left|
                 \begin{array}{cc}
                   \alpha(s,u) & \gamma(s,u) \\
                   \beta(s,u) & \delta(s,u) \\
                 \end{array}
               \right|
}{t\bar{v}-\bar{t}v}\\
&=(\alpha(s,u)\delta(s,u)-\beta(s,u)\gamma(s,u))\frac{\left|
          \begin{array}{cc}
            F(s,u;t,v) & G(s,u;t,v) \\
            F(s,u;\bar{t},\bar{v})  & G(s,u;\bar{t},\bar{v}) \\
          \end{array}
        \right|
}{t\bar{v}-\bar{t}v}.
\end{aligned}
\end{equation*}
Hence
\begin{equation}
B_{t,v}(\widetilde{F},\widetilde{G})=(\alpha(s,u)\delta(s,u)-\beta(s,u)\gamma(s,u))B_{t,v}(F,G),
\end{equation}
so
\begin{equation}
f_i(s,u)=(\alpha(s,u)\delta(s,u)-\beta(s,u)\gamma(s,u))\widetilde{f}_i(s,u).
\end{equation}
Now $(\alpha(s,u)\delta(s,u)-\beta(s,u)\gamma(s,u))=0$ if and only
if the two syzygies $\widetilde{\mathbf{p}},\widetilde{\mathbf{q}}$
are linear dependent. But by assumption
$$\gcd\big(\alpha(s,u)\delta(s,u)-\beta(s,u)\gamma(s,u),h(s,u)\big)=1,$$
 where $h(s,u)$ is the inversion formula for the
singularity $\mathbf{Q}$. Therefore
$$f_i^{\mathbf{Q}}=\widetilde{f}_i^{\mathbf{Q}}.$$\qed

\medskip

Hence to prove Theorem \ref{T:conjec_newversionforFG}, we  need to
focus only on the Smith normal form of the Bezout matrix
$B_{t,v}(\widetilde{F},\widetilde{G})$ constructed from another pair
of syzygies of the curve. Now suppose $\mathbf{Q}=(0,0,1)$ is an
order $r$ singularity on the curve $\mathbf{P}(s,u)$. Then the
degree $n$ curve $\mathbf{P}(s,u)$ has a parametrization:
\begin{equation}
\mathbf{P}(s,u)=(a(s,u)h(s,u),b(s,u)h(s,u),c(s,u)), \label{ahbhc}
\end{equation}
where $\gcd(a,b)=\gcd(h,c)=1$ and the roots of $h(s,u)$ are all the
parameters corresponding to the singularity $\mathbf{Q}$. Moreover,
we can perform a coordinate transformation so that $\gcd(a,h)=1$.
 As before, we first transfer all the information
for the singularity $\mathbf{Q}$ from $B_{t,v}(F,G)$ to
$B_{t,v}(M,L)$, where $M(s,u;t,v)$ and $L(s,u;t,v)$ are defined in
Equation (\ref{eq_ML}). Indeed by Theorem \ref{muabdsyzygy} we
 get the following result.

\begin{corollary}
$$S^{\mathbf{Q}}(B_{t,v}(F,G))=S^{\mathbf{Q}}(B_{t,v}(M,L)).$$
\label{C:FGML}
\end{corollary}

\subsection{The Smith normal forms of $B_{t,v}(\overline{M},L)$ and $B_{t,v}(h,L)$}
Since $M(s,u;t,v)=\overline{M}(s,u;t,v)h(t,v)$,  to study the Bezout
matrix $B_{t,v}(M,\\L)$, we shall next turn to the Smith normal
forms of the matrices $B_{t,v}(\overline{M},L)$ and $B_{t,v}(h,L)$.
Note that here we are switching from a Bezout matrix to two Hybrid
Bezout matrices because
$\deg_{t,v}(M)=\deg_{t,v}(L)=\deg_{t,v}(\overline{M})+\deg_{t,v}(h)$.

\medskip

 For brevity we shall assume that the singularity $\mathbf{Q}$
has ordinary infinitely near singularities only in its first
neighborhood. The more general cases can be treated similarly (see
Remark \ref{R:finalinduction}, below).

When we blow up the original curve $\mathbf{P}(s,u)$ in
(\ref{ahbhc}) (see \cite{JiaGoldman} for details), we get the curve
$$\mathbf{P}^1(s,u)=(a^2h,bc,ca).$$
Let $F^1(s,u;t,v)$ and $G^1(s,u;t,v)$ be the two algebraic curves
constructed from a $\mu$-basis for the new curve $\mathbf{P}^1(s,u)$
in the same way as we define $F(s,u;t,v)$ and $G(s,u;t,v)$ from a
$\mu$-basis for $\mathbf{P}(s,u)$. Note that
$\deg_{t,v}(F^1)=\deg_{t,v}(G^1)=\deg(\mathbf{P}^1)-1=2n-r-1$, where
$r$ is the order of the singularity $\mathbf{Q}$.

\begin{theorem}
\begin{equation}
\begin{aligned}
&S^{\mathbf{Q}}(B_{t,v}(F^1,G^1))\\&=\left(
    \begin{array}{ccccccc}
     1  &  &  &  &  &  &\\
       &\ddots  &  &  &  & & \\
       &  &1~~  &  &  &  &\\
       &  &  &\psi_r(s,u)  &  &  &\\
       &  &  && \psi_r(s,u)\psi_{r-1}(s,u)  &  &\\
       &  &  &  && \ddots  & \\
      &    &   &   &  & &\prod\limits_{i=2}^{r}\psi_i(s,u)
    \end{array}
  \right),
\end{aligned}
\end{equation}
where $\psi_i(s,u)$ are the inversion formulas for all the order $i$
infinitely near singularities of $\mathbf{Q}$. \label{bottom}
\end{theorem}

\noindent Proof. Let the Smith normal form of $B_{t,v}(F^1,G^1)$ be
$$\diag(f_{2n-r},f_{2n-r}f_{2n-r-1},\cdots,f_{2n-r}\cdots f_2).$$
 Then by a result similar to Corollary 4 in \cite{8Chen03} (See Theorem \ref{T:Chen_another version} in the Appendix) and Corollary \ref{sameinvariant},
\begin{equation}
\psi_i(s,u)|f_i(s,u),
\end{equation}
so
\begin{equation}
\psi_i(s,u)|f_i^{\mathbf{Q}}(s,u). \label{divides}
\end{equation}
Suppose that there are
 $m_i$ infinitely near
singularities
 of order $i$ related to
 the point $\mathbf{Q}$ on the new curve $\mathbf{P}^1(s,u)$. Then
  by Proposition \ref{intesectionmultiplicity for FG} and Equation
  (\ref{divides})
 \begin{equation}
 \begin{aligned}
 \sum\limits_{\mathbf{Q}^*}I_{\mathbf{Q}^*}(F^1,G^1)=\sum\limits_{i=1}^r m_i\times
 i\times(i-1)&=\sum\limits_{i=1}^r\deg(\psi_i)\times(i-1)\\
&\leq \sum\limits_{i=1}^r\deg(f_i^{\mathbf{Q}})\times(i-1),
 \label{h1}
 \end{aligned}
 \end{equation}
 where the sum is taken over all the order $i$ infinitely near singularities
 of $\mathbf{Q}$ not including $\mathbf{Q}$ itself.
On the other hand, since all the factors $f_i^{\mathbf{Q}}$
contribute to the intersection number $
\sum\limits_{\mathbf{Q}^*}I_{\mathbf{Q}^*}(F^1,G^1)$, and for each
$i$, the factor $f_i^{\mathbf{Q}}$ appears in the last $i-1$
positions of the Smith normal form of the matrix $B_{t,v}(F^1,G^1)$,
\begin{equation}
 \sum\limits_{\mathbf{Q}^*}I_{\mathbf{Q}^*}(F^1,G^1)\geq
 \sum\limits_{i=1}^r\deg(f_i^{\mathbf{Q}})\times (i-1),
 \label{h2}
\end{equation}
Hence Equation (\ref{h1}) and Equation (\ref{h2}) yield
$$\deg(f_i^{\mathbf{Q}})=\deg(\psi_i),~~~i=1,\cdots,r.$$
Therefore by Equation (\ref{divides})
$$f_i^{\mathbf{Q}}(s,u)=\psi_i(s,u),~~~i=1,\cdots,r.$$
 Also since the orders of the infinitely
near singularities of $\mathbf{Q}$ are less than or equal to the
order of $\mathbf{Q}$,
$$f_i^{\mathbf{Q}}(s,u)=1~~\hbox{for}~~i>r.$$\qed

We shall next transfer the information on the singularity
$\mathbf{Q}$ from the Smith normal form of $B_{t,v}(F^1,G^1)$ to the
Smith normal form of $B_{t,v}(\overline{M},L)$.

\begin{theorem}
$$S^{\mathbf{Q}}(B_{t,v}(\overline{M},L))=S^{\mathbf{Q}}(B_{t,v}(F^1,G^1))_{(n-1)\times(n-1)},$$
where the subscript $(n-1)\times(n-1)$ means the $(n-1)\times(n-1)$
submatrix in the  lower right corner.
 \label{bridge}
\end{theorem}
\noindent Proof. For the blow up curve
$$\mathbf{P}^1(s,u)=(a^2h,bc,ca),$$ we have a pair of syzygies
\begin{equation}
\mathbf{S}^1(s,u)\triangleq (0,a,-b),~~~\mathbf{T}^1(s,u)\triangleq
(c,0,-ah).
\end{equation}
Construct two polynomials from $\mathbf{S}^1(s,u)$ and
$\mathbf{T}^1(s,u)$:
\begin{equation}
\begin{aligned}
S_1(s,u;t,v)&\triangleq
\frac{\mathbf{S}_1(s,u)\cdot\mathbf{P}^1(t,v)}{sv-tu}\\
&=\frac{a(s,u)b(t,v)-b(s,u)a(t,v)}{sv-tu}c(t,v),\\
T_1(s,u;t,v)&\triangleq
\frac{\mathbf{T}_1(s,u)\cdot\mathbf{P}^1(t,v)}{sv-tu}\\
&=\frac{c(s,u)a(t,v)h(t,v)-c(t,v)a(s,u)h(s,u)}{sv-tu}a(t,v).
\end{aligned}
\label{S1T1}
\end{equation}
Since $\mathbf{S}^1,\mathbf{T}^1$ are a pair of syzygies for the
curve $\mathbf{P}^1(s,u)$, by Theorem \ref{muabdsyzygy}
$$S^{\mathbf{Q}}(B_{t,v}(S^1,T^1))=S^{\mathbf{Q}}(B_{t,v}(F^1,G^1)).$$
Comparing the expressions for $\overline{M}, L$ in Equation
(\ref{eq_ML}) with the expressions for $S^1,T^1$ in Equation
(\ref{S1T1}), and recalling that $\gcd(h,c)=\gcd(a,h)=1$, we
conclude that
$$S^{\mathbf{Q}}(B_{t,v}(\overline{M},L))=S^{\mathbf{Q}}(B_{t,v}(F^1,G^1)).$$\qed

\bigskip

From the previous two theorems we know that the Smith normal form of
the Bezout matrix $B_{t,v}(\overline{M},L)$ provides the parameters
for all the infinitely near singularities of the singular point
$\mathbf{Q}$. Next we shall show that the Smith normal form of
$B_{t,v}(h(t,v),L(s,u;t,v))$ provides all the parameters for the
singularity $\mathbf{Q}$ itself.

\begin{theorem}
\begin{equation}
S^{\mathbf{Q}}(B_{t,v}(h(t),L(s,u;t,v)))=\diag(\underbrace{1,\cdots,1}_{n-r}\underbrace{,h(s,u),\cdots,h(s,u)}_{r-1}).
\end{equation}
\label{h_intersection}
\end{theorem}
\noindent Proof. Denote by
$$B\triangleq B_{t,v}((sv-tu)h(t,v),(sv-tu)L).$$
By Lemma \ref{sameinvariant}, we only need to prove that
\begin{equation*}
\begin{aligned}
S^{\mathbf{Q}}(B)
=\diag(\underbrace{1,\cdots,1}_{n-r}\underbrace{,h(s,u),\cdots,h(s,u)}_{r-1},0).
\end{aligned}
\end{equation*}
Since \begin{equation*}
(sv-tu)L(s,u;t,v)=c(s,u)a(t,v)h(t,v)-c(t,v)a(s,u)h(s,u),
\end{equation*}
and the polynomials $c(s,u)a(t,v)h(t,v)$ and $c(t,v)a(s,u)h(s,u)$
are both of degree $n$ in $(t,v)$,
 we have $$B\approx B_1+B_2,$$ where
\begin{equation*}
\begin{aligned}
&B_1=c(s,u)B_{t,v}(h(t,v)(sv-tu),a(t,v)h(t,v)),\\&B_2=a(s,u)h(s,u)B_{t,v}(h(t,v)(sv-tu),c(t,v)),
\end{aligned}
\end{equation*}
and the notation $\approx$ means that the first $r+1$ rows of the
matrices $B$ and $B_1+B_2$ are the same, while the entries in the
last $n-r-1$ rows of the matrix $B_1+B_2$ are equal to twice the
corresponding entries in the matrix $B$. We can examine the Smith
normal form of the matrix $B_1+B_2$ instead of the Smith normal form
of the matrix $B$ because we are  interested only in the polynomials
in the Smith normal form, so the multiplication by two does not
matter.

\medskip

 Let
$H_{k}(s,u)$ be an order $k$ submatrix of the matrix $B_1+B_2$. Then
$H_{k}(s,u)=H_{k,1}(s,u)+H_{k,2}(s,u),$ where $H_{k,1}(s,u)$ and
$H_{k,2}(s,u)$ are order $k$ submatrices of $B_1$ and $B_2$, so by
\cite{S.J.Xu}
\begin{equation}
\det(H_k(s,u))=\det(H_{k,1}(s,u))+\sum\limits_{j=1}^{k-1}\Gamma_n^i\det(H_{1}/H_{2}^i)+\det(H_{k,2}(s,u)),
\label{sumdet}
\end{equation}
where $\Gamma_n^i\det(H_{1}/H_{2}^i)$ is the sum of the combination
of determinants in which  $i$ rows of $H_{1,k}$ are replaced by the
corresponding rows of the matrix $H_{k,2}$. Note that
$$\rank(B_1)=n-r$$ because
$\gcd(h(t,v)(sv-tu),a(t,v)h(t,v))=h(t,v)$.
 Hence
$$\det(H_{n-r+1,1}(s,u))\equiv 0.$$
 Also note that
every element in $B_2$ is a multiple of $a(s,u)h(s,u)$. Therefore
from Equation (\ref{sumdet}) we know that
$$a(s,u)h(s,u)|\det(H_{n-r+1}(s,u)).$$
Since the $k$-th determinant factor $D_k(s,u)$ of the matrix
$B_1+B_2$ is the GCD of the $k\times k$ minors of $B_1+B_2$,
\begin{equation}
a(s,u)h(s,u)|D_{n-r+1}(s,u). \label{ah|D}
\end{equation}
Suppose that the Smith normal form of the matrix $B_1+B_2$ is
$$S(B_1+B_2)=\diag(f_n, f_{n-1}, \cdots, f_2, f_1),$$
By Theorem \ref{T:Chen_another version} in the Appendix, for any
root $(s,u)$ of the polynomial $h$, $f_i(s,u)\not=0$ for $i>r$.
Hence since $f_{i+1}|f_i$, it follows by Equation (\ref{ah|D}) that
$$h(s,u)|f_i,~~i=1,\cdots,r.$$
Note that $\det(B)\equiv0$ because
$\gcd((sv-tu)h(t,v),(sv-tu)L)\not=1$. Moreover by Proposition
\ref{intersectionforh} the intersection multiplicity
$$I_{\mathbf{Q}}(h(t,v),L(s,u;t,v))=r(r-1)=\deg(h)\times(r-1),$$ so
\begin{equation*}
S^{\mathbf{Q}}(B)=\diag(\underbrace{1,\cdots,1}_{n-r}\underbrace{,h(s,u),\cdots,h(s,u)}_{r-1},0).
\end{equation*}
\qed

\medskip

Now we have computed the Smith normal forms of the Bezout matrices
$B_{t,v}(\overline{M},L)$ and $B_{t,v}(h,L)$. Since
$\det(B_{t,v}(M,L))=\Res_{t,v}(M,L)$,
$$\det(B_{t,v}(M,L))=\det(B_{t,v}(\overline{M},L))\det(B_{t,v}(h,L)).$$
Unfortunately
$$B_{t,v}(M,L)\not=B_{t,v}(\overline{M},L)B_{t,v}(h,L),$$ so
$$S(B_{t,v}(M,L))\not=S(B_{t,v}(\overline{M},L))S(B_{t,v}(h,L)),$$
 Therefore we
need some additional preparation to combine
$S(B_{t,v}(\overline{M},L))$ and $S(B_{t,v}(h,L))$ to get
$S(B_{t,v}(M,L))$.

\subsection{Companion matrices and factorization of Hybrid Bezout
matrices} In this subsection we shall introduce companion matrices
to factor Hybrid Bezout matrices and prepare for the later
recombination of $S(B_{t,v}(\overline{M},L))$ and $S(B_{t,v}(h,L))$.

\medskip

In this subsection $\mathbb{D}$ is an integral domain of
characteristic zero.
\begin{definition}
Let $P(t)$ be a degree $n$ polynomial in $\mathbb{D}[t]$:
$$P(t)=p_0t^n+p_1t^{n-1}+\cdots+p_n,~~~p_0\not=0.$$
The companion matrix of $P(t)$ is defined by:
$$\Delta_P=\left(
             \begin{array}{ccccc}
               0 & 0 & \cdots & 0 & -p_n \\
               p_0 & 0 & \cdots & 0 & -p_{n-1} \\
               0 & p_0 & \cdots & 0 & -p_{n-2} \\
               \vdots & \vdots & \ddots & \vdots & \vdots \\
               0 & 0 & \cdots & p_0 & -p_1 \\
             \end{array}
           \right).
$$
\end{definition}

The following proposition states the well known relationship between
companion matrices and the resultant of two univariate polynomials.
\begin{proposition}\cite{GAMA,Cox}
Let $P, Q$ be two polynomials in $\mathbb{D}[t]$ with $m=\deg(Q)\leq
\deg(P)=n$. Then
$$\Res(Q,P)=p_0^m \det(Q(\Delta_{P/p_0})),$$
where $Q(\Delta_{P/p_0})$ refers to the evaluation of the polynomial
$Q$ at the  matrix $\Delta_{P/p_0}$.
 \label{companionresultant}
\end{proposition}

\medskip

By Proposition \ref{companionresultant},
$\Res(QR,P)=\Res(Q,P)\Res(R,P).$ Generally, however,
$B(QR,P)\not=B(Q,P)B(R,P)$, but the following proposition provides a
resultant matrix which can be factored in this way.

\begin{proposition}\cite{GAMA}
Let $P,Q,R$ be polynomials in $\mathbb{D}[t]$ satisfying
$\deg(Q)+\deg(R)\leq\deg(P)$. Let
\begin{equation}
H(Q,P)\triangleq J_n \cdot Q(\Delta_{P/p_0}^t)\cdot J_n, \label{H}
\end{equation}
where
$$J_n=\left(
                    \begin{array}{ccc}
                       &  & 1 \\
                       & {\mathinner{\mkern2mu\raise1pt\hbox{.}\mkern2mu
\raise4pt\hbox{.}\mkern2mu\raise7pt\hbox{.}\mkern1mu}}
 &  \\
                      1 &  &  \\
                    \end{array}
                  \right)_{n \times n}.$$
 Then
$$H(QR,P)=H(Q,P)\cdot H(R,P).$$
\label{productlemma}
\end{proposition}
\noindent Proof.
\begin{equation*}
\begin{aligned}
H(Q,P)\cdot H(R,P)&=J_n \cdot Q(\Delta_{P/p_0}^t)\cdot J_n\cdot J_n
\cdot R(\Delta_{P/p_0}^t)\cdot J_n\\
&=J_n\cdot Q(\Delta_{P/p_0}^t)\cdot R(\Delta_{P/p_0}^t)\cdot J_n\\
&=J_n\cdot QR(\Delta_{P/p_0}^t)\cdot J_n\\
&=H(QR,P).
\end{aligned}
\end{equation*}\qed

 The following factorization shows the relationship
between Hybrid Bezout resultant matrices and the companion resultant
matrices defined in Proposition \ref{productlemma}.
\begin{proposition}\cite{GAMA}
Let $P,Q$ be two polynomials in $\mathbb{D}[t]$ with $m=\deg(Q)\leq
\deg(P)=n$. Let $B(Q,P)$ be the Hybrid Bezout resultant matrix of
$P$ and $Q$ with respect to $t$, and let $H(Q,P)$ be the matrix
defined in Equation (\ref{H}).  Then
$$B(Q,P)=T_m\cdot H(Q,P),$$
where
$$T_m=\left(
      \begin{array}{cccccc}
        p_0 & \cdots & p_{m-1} & 0 & \cdots & 0 \\
         & \ddots & \vdots & \vdots &  & \vdots \\
        &  & p_0 & 0 &  & 0 \\
        0 & \cdots & 0 & 1 &  &  \\
        \vdots &  & \vdots &  & \ddots &  \\
        0 & \cdots & 0 &  &  & 1 \\
      \end{array}
    \right)_{n\times n},~~
$$
and $Q(\Delta_{P/p_0}^t)$ refers to the evaluation of the polynomial
$Q$ at the transpose of the matrix $\Delta_{P/p_0}$.
\label{companion lemma}
 \end{proposition}

\begin{theorem}
Let $f,g,h$ be polynomials in $\mathbb{D}[t]$ with $\deg(f)=m,
\deg(g)=n$ and $\deg(h)\geq m+n$ . Denote by $\alpha_k, \beta_k, \gamma_k$
the $k$-th invariant factors of the Hybrid Bezout matrices
$B(f,h)$, $B(g,h)$ and $B(fg,h)$. Then
$$\alpha_{i_1}\alpha_{i_2}\cdots\alpha_{i_k}\beta_{j_1}\beta_{j_2}\cdots\beta_{j_k}|h_0^l\gamma_{i_1+j_1-1}\gamma_{i_2+j_2-2}\cdots\gamma_{i_k+j_k-k},,$$
where $h_0$ is the leading coefficient of the polynomial $h(t)$, and
$l$ is some non-negative integer.
 \label{elementary divisors lemma}
\end{theorem}
\noindent Proof. By Proposition \ref{companion lemma},
$$B(f,h)=T_m\cdot H(f,h),~B(g,h)=T_n\cdot H(g,h),$$
and $$B(fg,h)=T_{m+n}\cdot H(fg,h).$$ Since by Proposition
\ref{productlemma},
$$H(fg,h)=H(f,h)\cdot H(g,h),$$
we have
\begin{equation}
T_{m+n}\cdot T_m^{-1}\cdot B(f,h)\cdot T_n^{-1}\cdot B(g,h)=B(fg,h).
\label{Bezoutproduct}
\end{equation}
Note that now our equality holds over the $\mathbb{D}[t,h_0^{-1}]$.
The entries in the matrices $T_m^{-1}$ and $T_n^{-1}$ have
denominators $h_0^m$ and $h_0^n$, so multiplying both sides of
Equation (\ref{Bezoutproduct}) by $h_0^{m+n}$ to clear these
denominators yields
\begin{equation}
T_{m+n}\cdot (h_0^{m}T_m^{-1})\cdot B(f,h)\cdot (h_0^nT_n^{-1})\cdot
B(g,h)=h_0^{m+n}B(fg,h). \label{bezou2}
\end{equation}
Now the left hand side of Equation (\ref{bezou2}) is a product of
polynomial matrices. From Proposition \ref{P:invariants} we get directly get that
$$\alpha_{i_1}\alpha_{i_2}\cdots\alpha_{i_k}\beta_{j_1}\beta_{j_2}\cdots\beta_{j_k}|h_0^l\gamma_{i_1+j_1-1}\gamma_{i_2+j_2-2}\cdots\gamma_{i_k+j_k-k}$$
for some $l$.
\qed

\subsection{Joining  $S(B_{t,v}(\overline{M},L))$ and $S(B_{t,v}(h,L))$}

Now we are ready to join
$S(B_{t,v}(\overline{M}(s,u;t,v),L(s,u;t,v)))$ and
$S(B_{t,v}(h(t,v),\\
L(s,u;t,v)))$ together to compute
$S(B_{t,v}(M,L))$.

\begin{theorem}
\begin{equation*}
\begin{aligned}
&S^{\mathbf{Q}}(B_{t,v}(M,L))\\=&\diag(1,\cdots,1,h(s,u)\psi_r(s,u),h(s,u)\psi_r(s,u)\psi_{r-1}(s,u),\cdots,\\&
~~~~~~~~~~~~~~~~~~~~~~~~~~~~~~~~~~~~~~~~~~~~~~~~~~~~~~~~h(s,u)\prod\limits_{i=2}^{r}\psi_i(s,u)),
  \end{aligned}
\end{equation*}
where $\psi_i(s,u)$ is the inversion formula for all the order $i$
infinitely near singularities of $\mathbf{Q}$. \label{elementaryML}
\end{theorem}
\noindent Proof.
Denote the Smith normal form of $B_{t,v}(\overline{M},L)$ by
$$\diag(\overline{g}_1,\cdots,\overline{g}_{n-1}),$$
the Smith normal form of $B_{t,v}(h,L)$ by
$$\diag(\widetilde{g}_1,\cdots,\widetilde{g}_{n-1}),$$
and the Smith
 form of $B_{t,v}(M,L)$ by
$$\diag(g_1,\cdots,g_{n-1}).$$
 By Theorem \ref{bridge} we know that
\begin{equation}
\overline{g}_i^{\mathbf{Q}}=\left\{
                                  \begin{array}{ll}
                                    1, & \hbox{for}~~ 1\leq i< n-r+1 \\
                                     \prod\limits_{k=2}^{n-i+1}\psi_k(s,u), & \hbox{for} ~~n-r+1\leq i\leq n-1.
                                  \end{array}
                                \right.
                                \label{equ_of_Mbar}
\end{equation}
Also by Theorem \ref{h_intersection} we know that
\begin{equation}
\widetilde{g}_i^{\mathbf{Q}}=\left\{
                                  \begin{array}{ll}
                                    1, & \hbox{for}~~ 1\leq i< n-r+1 \\
                                    h(s,u), & \hbox{for} ~~n-r+1\leq i\leq n-1.
                                  \end{array}
                                \right.
                                \label{equ_of_h}
\end{equation}
By Proposition  \ref{P:invariants}, for $i=1,\cdots,r-1$ we have
\begin{equation}
\overline{g}_1\overline{g}_2\cdots\overline{g}_{n-r}\overline{g}_{n-r+1}\widetilde{g}_1\widetilde{g}_2\cdots\widetilde{g}_{n-r}\widetilde{g}_{n-r+i}|h_0^lg_1g_2\cdots g_{n-r}g_{n-r+i},
\end{equation}
for some non-negative integer $l$, where $l_0(s,u)$ is the leading
coefficient of the polynomial $L(s,u;t,v)$ in $(t,v)$.
But
$$l_0(s,u)=lc(ah)c(s,u)-lc(c)a(s,u)h(s,u),$$ where $lc$ means the
leading coefficient of the polynomial. Thus
$$\gcd(l_0(s,u),h(s,u))=1.$$ Therefore, when restricted to the point
$\mathbf{Q}$, we  have
\begin{equation}
\overline{g}_1\overline{g}_2\cdots\overline{g}_{n-r}\overline{g}_{n-r+1}\widetilde{g}_1\widetilde{g}_2\cdots\widetilde{g}_{n-r}\widetilde{g}_{n-r+i}|g_1g_2\cdots g_{n-r}g_{n-r+i},
\end{equation}
which by (\ref{equ_of_Mbar}) and (\ref{equ_of_h}) is equivalent to
\begin{equation}
\psi_2\cdots\psi_{r-i+1}h|g_{n-r+i}.
\end{equation}
Since $\det(B_{t,v}(\overline{M},L))\det(B_{t,v}(h,L))=B_{t,v}(M,L)$,
we immediately get
\begin{equation}
\psi_2\cdots\psi_{r-i+1}h=g_{n-r+i},~i=1,\cdots,r-1.
\end{equation}
up to a constant multiple. The proof is then complete.\qed

\bigskip

By Theorem \ref{elementaryML} and Corollary \ref{C:FGML}, for all
the singularities on the curve $\mathbf{P}(s,u)$, we finally have
$$d_k(s,u)=\prod\limits_{\mathbf{Q}}d_k^{\mathbf{Q}}=h_k(s,u)\prod\limits_{i\geq k}\psi_r^i(s,u).$$
Theorem \ref{T:conjec_newversionforFG} is now proved.

\begin{figure}[!htb]
\centering
         \includegraphics[width=4in]{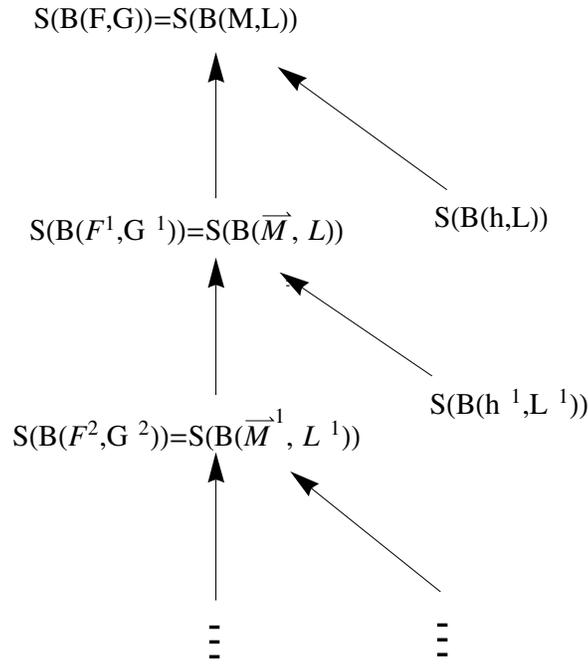}
\caption{Proof by induction on the height of the singularity tree.}
\label{induction}
\end{figure}

\begin{remark}
\label{R:finalinduction} At the beginning of Section 5.2, we assume
that all the singularities on the original curve $\mathbf{P}(s,u)$
can be totally resolved after one blow-up of the curve. Actually our
proof works in general where an arbitrary number of $k$ blow-ups are
needed to totally resolve all the singularities on the curve
$\mathbf{P}(s,u)$. The proof is by induction. Suppose that a
singularity $\mathbf{Q}$ on the curve $\mathbf{P}(s,u)$ has
infinitely near singularities in the $k$-th neighborhood. We start
from the $k$-th blow-up curve, whose singularities have no
infinitely near singularities. Then by our proof, the Smith normal
form of the Bezout matrix of the two polynomials $F^{k-1}$ and
$G^{k-1}$ constructed from a $\mu$-basis for the $k-1$-st blow-up
curve contains all the singularities on the $k$-th blow-up curve
(given by $S(B(\overline{M}^{k-1},L^{k-1}))$) together with all the
basic singularities on the $k-1$-st blow up curve itself (given by
$S(B(h^{k-1},L^{k-1}))$). We can continue with this method
proceeding by induction until we reach the top of the singularity
tree. (See Figure \ref{induction}).
\end{remark}

\section{A Discussion of the Conjecture}
Let $\mathbf{P}(t,v)=(a(t,v),b(t,v),c(t,v))$ be a rational planar
curve with a $\mu$-basis $\mathbf{p}(s,u),\mathbf{q}(s,u)$, and let
$\mathbf{L}_1(s,u)=(c(s,u),0,-a(s,u)),
\mathbf{L}_2(s,u)=(0,c(s,u),\\-b(s,u))$ be a pair of obvious
syzygies of the curve $\mathbf{P}(t,v)$. Then to compute all the
singularities on the curve $\mathbf{P}(t,v)$, we can compute the
Smith normal form for any one of the following four Bezout resultant
matrices.

\begin{enumerate}
  \item $B_{s,u}(\mathbf{p}(s,u)\cdot\mathbf{P}(t,v),\mathbf{q}(s,u)\cdot\mathbf{P}(t,v))$
  \item $B_{t,v}(\mathbf{p}(s,u)\cdot\mathbf{P}(t,v),\mathbf{q}(s,u)\cdot\mathbf{P}(t,v))$
  \item $B_{s,u}(\mathbf{L}_1(s,u)\cdot\mathbf{P}(t,v),\mathbf{L}_2(s,u)\cdot\mathbf{P}(t,v))$
  \item $B_{t,v}(\mathbf{L}_1(s,u)\cdot\mathbf{P}(t,v),\mathbf{L}_2(s,u)\cdot\mathbf{P}(t,v))$
\end{enumerate}

Note that  only matrix 1 which is the focus of the conjecture of
Chen et al. is a Hybrid Bezout matrix, while the other three
matrices are Bezout matrices. Theoretically, it is easier to study
Bezout matrices than Hybrid Bezout matrices --- in fact, one of the
main obstructions to proving the conjecture of Chen, Wang and Liu is
to prove an analogue of Theorem \ref{muabdsyzygy} for Hybrid Bezout
matrices; which is why we study the Bezout matrix in 2 rather than
the Hybrid Bezout matrix in the conjecture.

\medskip

The Smith normal forms of these four  Bezout matrices are:
\begin{enumerate}
  \item $\diag(d_{n-\mu}(t,v),d_{n-\mu}(t,v)d_{n-\mu-1}(t,v),\cdots,d_{n-\mu}(t,v)\cdots
  d_2(t,v),0)$

  \item $\diag(1,\cdots,1,d_{n-\mu}(s,u),d_{n-\mu}(s,u)d_{n-\mu-1}(s,u),\cdots,\\~~~~~~~~~~~~~~~~~~~~~~~~~~~~~~~~~~~~~~~~~~~~~~
  ~~~~~~~~~~~~~d_{n-\mu}(s,u)\cdots d_2(s,u),0)$
   \item $c(s,u)\diag(1,\cdots,1,d_{n-\mu}(s,u),d_{n-\mu}(s,u)d_{n-\mu-1}(s,u),\cdots,\\~~~~~~~~~~~~~~~~~~~~~~~~~~~~~~~~~~~~~~~~~~~
  ~~~~~~~~~~~~~~~~d_{n-\mu}(s,u)\cdots d_2(s,u),0)$
  \item $c(s,u)\diag(1,\cdots,1,d_{n-\mu}(s,u),d_{n-\mu}(s,u)d_{n-\mu-1}(s,u),\cdots,\\~~~~~~~~~~~~~~~~~~~~~~~~~~~~~~~~~~~~~~~~~~~
  ~~~~~~~~~~~~~~~~d_{n-\mu}(s,u)\cdots d_2(s,u),0)$.
\end{enumerate}
The conjecture of Chen, Wang and Liu deals with the Hybrid Bezout
matrix of $F(s,u;t,v)$ and $G(s,u;t,v)$ with respect to the
parameter $(s,u)$, while what we actually proved is a result closely
related this conjecture dealing with the larger  Bezout matrix of
$F(s,u;t,v)$ and $G(s,u;t,v)$ with respect to the parameter $(t,v)$.
Geometrically, since by Proposition \ref{P:symmetric} the parameter
pair $(s,u;t,v)$ is an intersection point of the two algebraic
curves $F(s,u;t,v)=0$ and $G(s,u;t,v)=0$ if and only if the
parameter pair $(t,v;s,u)$ is also an intersection point of the two
algebraic curve $F(s,u;t,v)=0$ and $G(s,u;t,v)=0$, taking the Hybrid
Bezout matrix of $F(s,u;t,v)$ and $G(s,u;t,v)$ with respect to
parameter $(s,u)$ or taking the Bezout matrix of $F(s,u;t,v)$ and
$G(s,u;t,v)$ with respect to $(t,v)$ should give the same Smith
normal form except that the latter matrix has larger size. However,
currently we lack a rigorous algebraic proof for the equivalence of
these two Smith normal forms. Also  note that  although the Hybrid
Bezout matrix in the conjecture of Chen et al. is smaller than the
Bezout matrix in our main theorem --- $(n-\mu)\times (n-\mu)$ vs.
$n\times n$ --- the entries in our Bezout matrix with respect to
$(t,v)$ are lower degree polynomials than the polynomial entries in
the Hybrid Bezout matrix with respect to $(s,u)$ of Chen et al. ---
degree $n-\mu$ vs. degree $n$.

\medskip

We can also compute all the singularities of the curve
$\mathbf{P}(t,v)$ from the Smith normal forms of the Bezout matrices
of $\mathbf{L}_1(s,u)\cdot\mathbf{P}(t,v)$ and
$\mathbf{L}_2(s,u)\cdot\mathbf{P}(t,v)$ either with respect to the
parameter $(s,u)$ or with respect to the parameter $(t,v)$. Note
that both
$\mathbf{L}_1(s,u)\cdot\mathbf{P}(t,v)=c(s,u)a(t,v)-a(s,u)c(t,v)$
and
$\mathbf{L}_2(s,u)\cdot\mathbf{P}(t,v)=c(s,u)b(t,v)-b(s,u)c(t,v)$
are antisymmetric with respect to the parameters $(s,u)$ and
$(t,v)$. Therefore the Smith normal forms 3 and 4 are the same up to
a sign. Here, however, we need to remove the extra factor $c(s,u)$
or $c(t,v)$ from the Smith normal forms 3 or 4 to get the true
singularities of the curve $\mathbf{P}(t,v)$ because for any root
$(s^*,u^*)$ of the polynomial $c(s,u)$,
$\gcd(\mathbf{L}_1(s^*,u^*)\cdot\mathbf{P}(t,v),\mathbf{L}_2(s^*,u^*)\cdot\mathbf{P}(t,v))=c(t,v)$.

\section*{Acknowledgement} We would like to thank Laurent Buse for his help in pointing us to the companion matrices
used in Section 5.3. This work was partially supported by NSF grant
CCR-020331, by the 111 Project of China grant b07033, by the NSF of
China grant 10671192, and by the 100 Talent Project sponsored by CAS
of China.

\newpage

\hspace{4cm}Appendix {
\appendix
\makeatletter
\renewcommand\thetheorem{A.\@arabic\c@theorem}
\makeatother \setcounter{theorem}{0} \makeatletter
\renewcommand\theequation{E.\@arabic\c@equation}
\makeatother \setcounter{equation}{0}

Let $\mathbf{P}(t,v)$ be a rational planar curve with a $\mu$-basis
$\mathbf{p}(s,u),\mathbf{q}(s,u)$. Suppose that all the
singularities on the rational planar curve $\mathbf{P}(t,v)$ have no
infinitely near singularities. Then we have the following result
closely related to Corollary 4 in [Chen, Wang and Liu] which can be
derived from a very similar approach.
\begin{theorem}
The Smith normal form of the Bezout matrix
$B_{t,v}(\mathbf{p}(s,u)\cdot\mathbf{P}(t,v),\mathbf{q}(s,u)\cdot\mathbf{P}(t,v))$
is
$$\diag(1,\cdots,1,h_{n-\mu}(s,u),h_{n-\mu}h_{n-\mu-1},\cdots,h_{n-\mu}\cdots
h_2(s,u),0),$$ where $h_i(s,u)$ are the products of the inversion
formulas of all the order $i$ singularities on the curve
$\mathbf{P}(t,v)$. \label{T:Chen_another version}
\end{theorem}

To prove Theorem \ref{T:Chen_another version}, we shall prepare with
the following theorems. One can compare the outline of our proof
with the proofs from Lemma 2 to  Theorem 5 in [Chen, Wang and Liu].

\begin{theorem}
Let $\mathbf{Q}=\mathbf{P}(s_0,u_0)$ be an order $r$ singularity on
the curve $\mathbf{P}(s,u)$. Then
$$h(s,u)=\gcd(\mathbf{p}(s_0,u_0)\cdot\mathbf{P}(s,u),\mathbf{q}(s_0,u_0)\cdot\mathbf{P}(s,u))$$
is an inversion formula for the point $\mathbf{Q}$.
\label{T:anotherinversionfoumula}
\end{theorem}
\noindent Proof. Without loss of generality, we can assume that
$\mathbf{Q}=(0,0,1)$. Let
$$\mathbf{p}(s,u)=(p_1(s,u),p_2(s,u),p_3(s,u)),~~\mathbf{q}(s,u)=(q_1(s,u),q_2(s,u),q_3(s,u)).$$
Then
\begin{equation}
\begin{aligned}
&p_3(s_0,u_0)=\mathbf{p}(s_0,u_0)\cdot\mathbf{Q}=\mathbf{p}(s_0,u_0)\cdot\mathbf{P}(s_0,u_0)=0,\\
&q_3(s_0,u_0)=\mathbf{q}(s_0,u_0)\cdot\mathbf{Q}=\mathbf{q}(s_0,u_0)\cdot\mathbf{P}(s_0,u_0)=0.
\end{aligned}\label{E:app_h}
\end{equation}
 Since $\mathbf{Q}=(0,0,1)$,  the curve $\mathbf{P}(s,u)$ has
the parametrization
$$\mathbf{P}(s,u)=(a(s,u)h(s,u),b(s,u)h(s,u),c(s,u)),$$ where $\gcd(a,b)=\gcd(h,c)=1$.
Hence by Equation (\ref{E:app_h})
\begin{equation}
\begin{aligned}
&\gcd(\mathbf{p}(s_0,u_0)\cdot\mathbf{P}(s,u),\mathbf{q}(s_0,u_0)\cdot\mathbf{P}(s,u))\\
=&\gcd(p_1(s_0,u_0)ah+p_2(s_0,u_0)bh+p_3(s_0,u_0)c,\\
&
q_1(s_0,u_0)ah+q_2(s_0,u_0)bh+q_3(s_0,u_0)c)\\
=&\gcd(p_1(s_0,u_0)ah+p_2(s_0,u_0)bh,
q_1(s_0,u_0)ah+q_2(s_0,u_0)bh)\\
=&kh~~\text{for some polynomial}~ k.
\end{aligned}
\label{E:gcds}
\end{equation}
We claim that $k$ is a constant. Otherwise suppose that $(s^*,u^*)$
is a root of $k$. Then
\begin{equation*}
\begin{aligned}
&p_1(s_0,u_0)a(s^*,u^*)+p_2(s_0,u_0)b(s^*,u^*)=0\\
&q_1(s_0,u_0)a(s^*,u^*)+q_2(s_0,u_0)b(s^*,u^*)=0.
\end{aligned}
\end{equation*}
This means that the two vectors $\mathbf{p}(s_0,u_0)$ and
$\mathbf{q}(s_0,u_0)$ are linearly dependent, which is impossible.
Hence $k$ is a constant. Therefore, up to a constant multiple
$$h(s,u)=\gcd(\mathbf{p}(s_0,u_0)\cdot\mathbf{P}(s,u),\mathbf{q}(s_0,u_0)\cdot\mathbf{P}(s,u)).$$
 \qed

\bigskip

In the following theorems we denote the Bezout matrix
$B_{t,v}(\mathbf{p}(s,u)\cdot\mathbf{P}(t,v),\mathbf{q}(s,u)\cdot\mathbf{P}(t,v))$
by $B(s,u)$.
\begin{theorem}
The point $\mathbf{P}(s_0,u_0)$ is an order $r$ singular point if
and only if $\rank(B(s_0,u_0))=n-r$, where $r=\deg(\mathbf{P})$.
\label{T:rankloss}
\end{theorem}
\noindent Proof. By Theorem \ref{T:anotherinversionfoumula}, the
point $\mathbf{P}(s_0,u_0)$ is an order $r$ singular point if and
only if
$\deg(\gcd(\mathbf{p}(s_0,u_0)\cdot\mathbf{P}(s,u),\mathbf{q}(s_0,u_0)\cdot\mathbf{P}(s,u)))=r$.
By the standard properties of Bezout resultant matrices, the degree
of this gcd is $r$ if and only if $\rank(B(s_0,u_0))=n-r$.\qed

\bigskip

Recall that the order $k$ determinant factor of a matrix is the GCD
of all the order $k$ minors of the matrix. Let $D_k$ be the
determinant factors of the matrix $B(s,u),~k=1,\cdots,n.$
\begin{theorem}
Let $h(s,u)$ be an inversion formula of an order $r$ singular point
$\mathbf{Q}$ on the curve $\mathbf{P}(t,v)$. Then
$h(s,u)|D_{n-r+1}$. \label{T:divides}
\end{theorem}
\noindent Proof. Without loss of generality, we can assume that
$\mathbf{Q}=(0,0,1)$. Let
$$\mathbf{p}(s,u)=(p_1(s,u),p_2(s,u),p_3(s,u)),~~\mathbf{q}(s,u)=(q_1(s,u),q_2(s,u),q_3(s,u)).$$
Then by Remark \ref{R:inversionform},
\begin{equation}
\gcd(p_3,q_3)=\gcd(\mathbf{p}(s,u)\cdot\mathbf{Q},\mathbf{q}(s,u)\cdot\mathbf{Q})=h(s,u).
\label{E:gcd(p3,q3)}
\end{equation}%
Now let
$$\mathbf{P}(t,v)\triangleq\sum\limits_{i=0}^n(\lambda_{1i},\lambda_{2i},\lambda_{3i})t^iv^{n-1},$$
where $(\lambda_{1i},\lambda_{2i},\lambda_{3i})$ are constant
vectors. Then
\begin{equation*}
\begin{aligned}
&\mathbf{p}(s,u)\cdot\mathbf{P}(t,v)=\sum\limits_{i=0}^n(\lambda_{1i}p_1+\lambda_{2i}p_2+\lambda_{3i}p_3)t^iv^{n-i}\triangleq
\sum\limits_{i=0}^n\alpha_i(s,u)t^iv^{n-i}\\
&\mathbf{q}(s,u)\cdot\mathbf{P}(t,v)=\sum\limits_{i=0}^n(\lambda_{1i}q_1+\lambda_{2i}q_2+\lambda_{3i}q_3)t^iv^{n-i}\triangleq
\sum\limits_{i=0}^n\beta_i(s,u)t^iv^{n-i}.
\end{aligned}
\end{equation*}
By the construction of the Bezout matrices, the elements $b_{ij}$ in
the Bezout matrix
$B_{t,v}(\mathbf{p}(s,u)\cdot\mathbf{P}(t,v),\mathbf{q}(s,u)\cdot\mathbf{P}(t,v))$
are \cite{Fuhrmann}:
\begin{equation}
b_{ij}=\sum\limits_{k=1}^{m_{ij}}\alpha_{j+k-1}\beta_{i-k}-\alpha_{i-k}\beta_{j+k-1},
\label{E:elimentsinBezout}
\end{equation}
where $m_{ij}=\min\{i,n+1-j\}$. Therefore by Equations
(\ref{E:gcd(p3,q3)}) and (\ref{E:elimentsinBezout}), the Bezout
matrix $B(s,u)$ can be written as
\begin{equation}
B(s,u)=h(s,u)G(s,u)+H(s,u), \label{E:AppB}
\end{equation}
where $G(s,u)$ and $H(s,u)$ are matrices of size $n\times n$, and
 $H(s,u)$ has the form
\begin{equation}
H(s,u)=(q_2(s,u)p_1(s,u)-p_2(s,u)q_1(s,u))H_{0}, \label{E:appH}
\end{equation}
where $H_0$ is a constant matrix.
 Next we shall examine the rank of
$H(s,u)$. To do this we need to examine the rank of the constant
matrix $H_0$.

Let $(s_0,u_0)$ be a root of $h(s,u)$. Then by Theorem
\ref{T:rankloss} and Equation (\ref{E:AppB}),
\begin{equation}
\rank(H(s_0,u_0))=\rank(B(s_0,u_0))=n-r. \label{E:App_rank}
\end{equation}
 Since the $\mu$-basis
elements $\mathbf{p}(s_0,u_0)=(p_1(s_0,u_0),p_2(s_0,u_0),0)$ and
$\mathbf{q}(s_0,u_0)=(q_1(s_0,u_0),q_2(s_0,u_0),0)$ are linearly
independent,
$$q_2(s_0,u_0)p_1(s_0,u_0)-p_2(s_0,u_0)q_1(s_0,u_0)\not=0.$$ Hence
Equation (\ref{E:App_rank}) and Equation (\ref{E:appH}) yield
$\rank(H_0)=n-r$. Therefore, the polynomial matrix $H(s,u)$ has rank
$n-r$.

\medskip

 Let $B_{n-r+1}(s,u)$ be a size
$n-r+1$ submatrix of $B(s,u)$. Then
$$B_{n-r+1}(s,u)=h(s,u)G_{n-r+1}(s,u)+H_{n-r+1}(s,u).$$
Therefore,
$$\det(B_{n-r+1})=h^{n-r+1}\det(G_{n-r+1})+\cdots+\det(H_{n-r+1}).$$
Since $\rank(H(s,u))\equiv n-r$, $\det(H_{n-r+1})\equiv0$.
Therefore, $h|\det(B_{n-r+1})$. Hence $h|D_{n-r+1}$.\qed

\bigskip

Once we have Theorem \ref{T:divides}, we can apply the same approach
as in the rest of proofs of [Chen, Wang and Liu] to derive Theorem
\ref{T:Chen_another version}.}

\end{document}